%% file: HALIMI_Arxiv_2016_AltimSSE.tex
\documentclass[12pt,journal,draftcls,a4paper,onecolumn]{IEEEtran}

\usepackage[nomarkers,nofiglist,notablist]{endfloat} 

\usepackage{dsfont}
\usepackage{amsmath}
\usepackage{amssymb}
\usepackage{amsfonts}
\usepackage{graphicx}
\usepackage{amsmath}
\usepackage[all,poly]{xy}
\usepackage{multirow}
\usepackage{algorithm}
\usepackage{algorithmic}
\usepackage{color}
\usepackage[noadjust]{cite}
\usepackage{subfigure}
\usepackage{tikz,pgf}
\usepackage{url}
\usetikzlibrary{arrows,automata}
\usepackage{verbatim}
\usetikzlibrary{positioning,shapes.geometric}

\newcommand{\figwidth}{\columnwidth}

\newcommand{\bzero}{\boldsymbol{0}}

\newcommand{\beps}{\boldsymbol{\epsilon}}
\newcommand{\bthe}{\boldsymbol{\theta}}
\newcommand{\bThe}{\boldsymbol{\Theta}}
\newcommand{\bsig}{\boldsymbol{\sigma}}
\newcommand{\bSig}{\boldsymbol{\Sigma}}

\newcommand{\bGam}{\boldsymbol{\Gamma}}

\input notations.tex

\title{Bayesian Filtering of Smooth Signals: Application to Altimetry}

\author{Abderrahim Halimi$^{(1)}$\thanks{$^{(1)}$ Heriot-Watt University, School of Engineering and Physical Sciences, Edinburgh, U.K (e-mail: \{A.Halimi,G.S.Buller,s.mclaughlin\}@hw.ac.uk).}, Gerald S. Buller$^{(1)}$, Steve McLaughlin$^{(1)}$,  Paul Honeine$^{(2)}$\thanks{$^{(2)}$ The LITIS lab, Universit\'e de Rouen, Rouen, France (e-mail: paul.honeine@univ-rouen.fr)} \thanks{This work was supported by the EPSRC Grants EP/JO15180/1, EP/N003446/1, and EP/K015338/1} }

\begin{document}

\maketitle

\begin{abstract} 
This paper presents a novel Bayesian strategy for the estimation of smooth signals corrupted by Gaussian noise. The method assumes a smooth evolution of a succession of continuous signals that can have a numerical or an analytical expression with respect to some parameters. 
The Bayesian model proposed takes into account the Gaussian properties of the noise and the smooth evolution of the successive  signals. In addition, a gamma Markov random field prior is assigned to the signal energies and to the noise variances to account for their known properties. The resulting posterior distribution is maximized using a fast coordinate descent algorithm whose parameters are updated by analytical expressions. The proposed algorithm is tested on satellite altimetric data demonstrating good denoising results on both synthetic and real signals. The proposed algorithm is also shown to improve the quality of the altimetric parameters when combined with a parameter estimation strategy. 
\end{abstract}

%
%
%
%
%
%

\section{Introduction} \label{sec:Introduction}

%
%
%
%
%
%

\begin{figure}[h!]
\centering
\includegraphics[width=0.75\figwidth]{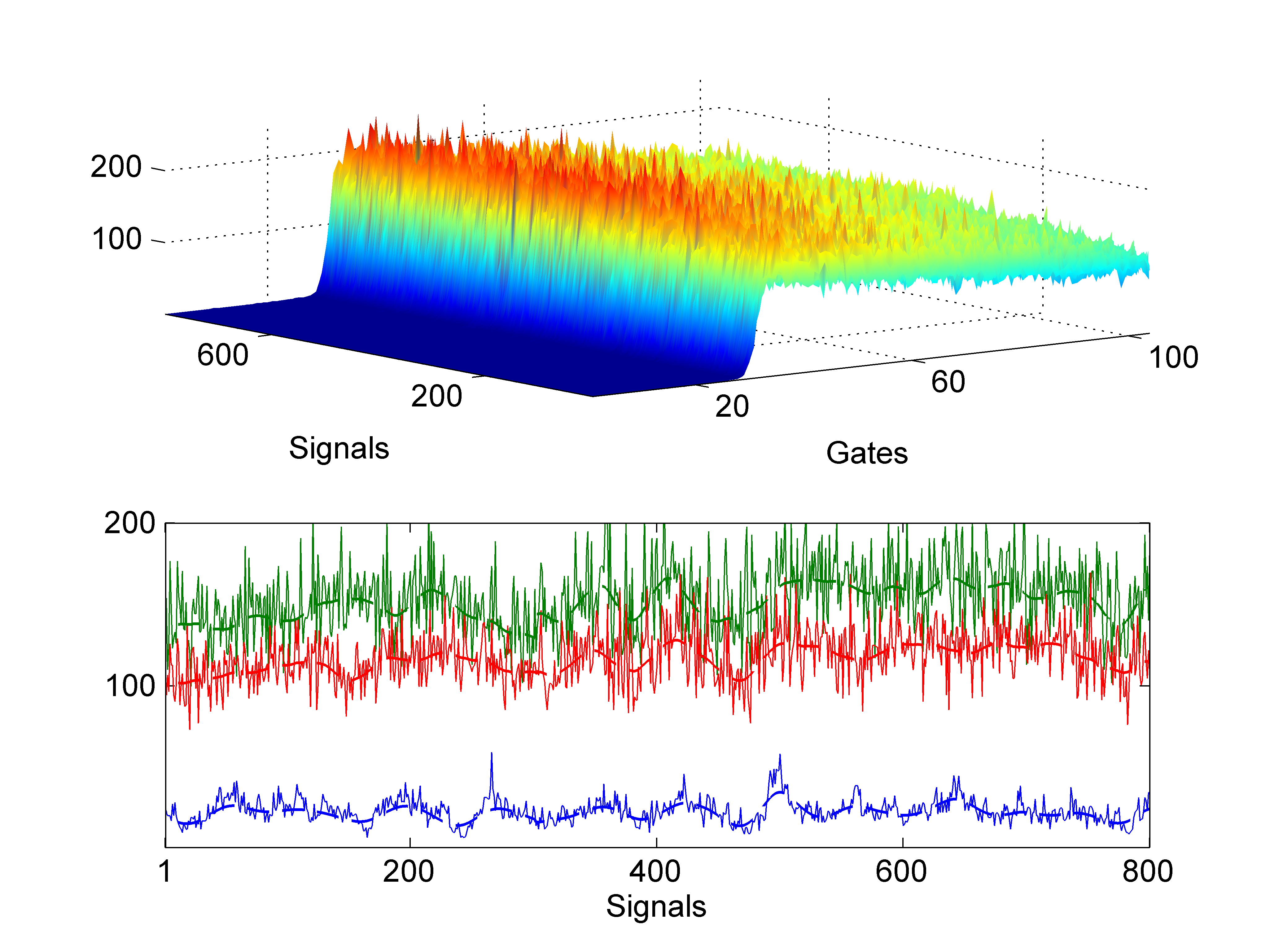}
\caption{(Top) Example of $800$ noisy Jason-$2$ signals. (Bottom) Signal evolutions in gates $(30,50,90)$ (continuous lines)  and their smooth approximation with the proposed SSE algorithm  (dashed lines).  } \label{fig:Real_Jason2_Echoes_JustifFilt}
\end{figure}

In many applications, the development of new sensor technologies allows for high speed acquisition of a succession of signals leading to a slight variation from one signal to the next. This is the case for  satellite altimetric signals that can be described as a succession of continuous functions corrupted by noise \cite{Hapke1981,Brown1977,HalimiTGRS2012}. Indeed, when observing the ocean, the altimetric successive signals show a reduced variation due to the nature of ocean (see Fig. \ref{fig:Real_Jason2_Echoes_JustifFilt} that shows a succession of $800$ signals acquired by the Jason-$2$  mission). This paper aims to exploit this correlation to denoise the observed altimetric signals.

A satellite altimeter is a nadir-viewing radar that emits regular pulses and records the travel time, the magnitude and the shape of each return signal after reflection on the Earth's surface. This reflected echo provides information about some physical parameters such as the range between the satellite and the observed scene (denoted by $\tau$), the significant wave height (denoted by SWH) and the wind speed (related to the signal's amplitude $P_u$). However, altimetric signals are corrupted by speckle noise and many recent studies and missions have been focusing on improving the quality of these signals by reducing the noise effect. This goal is generally achieved by considering two main approaches. The first approach improves the altimetric technology by increasing the number of observations (as for the Altika mission \cite{Vincent2006}) or by using a new delay/Doppler processing \cite{Raney1998}. The second approach improves the processing of the observed signals by considering more sophisticated physical models \cite{Brown1977,Amarouche2004,HalimiTGRS2012,Halimi2013}, or improved signal processing algorithms  \cite{Maus1998,Sandwell2005}. 
 This paper focuses on   signal processing approaches that can be divided into two categories. The first operates on the estimation algorithm to incorporate the known smooth properties of the altimetric parameters \cite{Maus1998,Sandwell2005,HalimiTGRS2015} while the second  operates on the observed signals to reduce the effects of noise   \cite{Ollivier2006,Thibaut2009OSTST}. This latter approach will be considered in this paper, i.e., noise reduction in the observed altimetric signals. The main motivation for this choice is to propose a denoising algorithm that is independent from the parameter estimation algorithm, thus, it can be easily combined with any existing estimation algorithms \cite{Amarouche2004,Maus1998,Sandwell2005,HalimiTGRS2015} leading to an improvement in parameter estimation. 

The first contribution of this paper is a hierarchical Bayesian model to denoise a set of smooth signals. Each signal is assumed  corrupted by additive, independent and non-identically distributed Gaussian noise.  This noise model generalizes the independent and identically distributed (i.i.d.) Gaussian noise that is generally assumed when considering altimetric data \cite{Amarouche2004,HalimiTGRS2014}. A gamma Markov random field (GMRF) prior \cite{DikmenTASLP2010} is considered to account for the correlation between the noise variances to better approximate the speckle noise. The signal energies are also assigned a GMRF prior to better approximate their continuity. Using Bayes rule, the likelihood and the prior distributions lead to a posterior distribution that will be used to estimate the noiseless signals and the noise parameters (as described in the next paragraph).  
Note that the proposed Bayesian hierarchy is generic in the sense that it does not assume a specific signal model. Indeed, the signal can be expressed by a numerical formula or given by linear/nonlinear analytical function with respect to (w.r.t.) some parameters.

The second contribution of this paper is the derivation of a denoising algorithm associated with the proposed hierarchical Bayesian model. The minimum mean square error (MMSE) and maximum a posteriori (MAP) estimators of the unknown signals/parameters cannot be easily computed from the obtained joint posterior. In this paper, the MAP estimator is evaluated by considering a coordinate descent algorithm (CDA) \cite{Bertsekas1995,Sigurdsson2014,HalimiTGRS2015} whose  convergence to a stationary point is ensured. The proposed algorithm sequentially updates the estimated noiseless signals,  noise variances and other hyperparameters by analytical formulas leading to a reduced computational cost. The proposed Bayesian model and estimation algorithm are validated using synthetic and real altimetric data acquired during the Jason-$2$ mission. The obtained results are very promising and show the potential of the proposed denoising strategy.

The paper is organized as follows. Section
\ref{sec:Problem_formulation} introduces the observation model and the considered altimetric signal. The proposed hierarchical Bayesian model and its estimation algorithm are introduced in Sections \ref{sec:Hierarchical_Bayesian_model} and \ref{sec:Estimation_algorithms}. Section \ref{sec:Validation_on_synthetic_data} validates the proposed technique using simulated data with controlled ground truth. Section \ref{sec:Results_on_Jason2_real_data} shows results obtained using real data resulting from the Jason-2 mission. Finally, 
conclusions and future work are  reported in Section
\ref{sec:Conclusions}.

\newpage
\section{Problem formulation} \label{sec:Problem_formulation}
Consider $M$ successive signals $\bsS \in \mathds{R}^{K\times M}$ and let $\bsY \in \mathds{R}^{K\times M}$  denote their noisy version. Let $\bsy_{:m} \in \mathds{R}^{K\times 1}$  be the $m$th column of $\bsY$ and $\bsy_{k:} \in \mathds{R}^{1\times M}$  its $k$th row, representing the $k$th temporal gate for all signals. For notation simplicity, we denote $\bsy_{:m}=\bsy_{m}$, for
$m = 1,\cdots, M$ and $\bsy_{k:}=\bsy_{k}$, for
$k = 1,\cdots, K$ (the same notation is used for $\bss$). The observation model is given by 
\begin{equation}
\bsy_m = \bss_m\left( \bThe_{m} \right) + \bse_m, \textrm{  with   } \bse_m \sim \calN \left(\bzero_K, \bSig \right)
\label{eqt:Observation_model}
\end{equation}
where $\sim$ means ``is distributed according to'', $\bsy_m$ and $\bss_m$ are  ($K \times 1$) vectors representing the $m$th observed and noiseless signals, and $\bse_m$ is a centered Gaussian noise vector with a diagonal covariance matrix $\bSig = \textrm{diag} \left(\bsig^2 \right)$ with $\bsig^2 =\left(\sigma^2_{1},\cdots,\sigma^2_{K}\right)^T$ a ($K\times 1$) vector. The signals $\bsS$ might  depend on some parameters  (by a linear or nonlinear expression) which are denoted by the ($1 \times H$) vector $\bThe_{m}=\left[\theta_1(m),\cdots,\theta_H(m)\right]$ containing the $H$  parameters of the $m$th signal. Note, however, that the proposed  method does not necessarily require a parametric expression for $\bsS$, and is valid provided that the signals satisfy some properties (as described in the following).
In different applications such as oceanic altimetry \cite{Maus1998,Sandwell2005}, the successive signals show a reduced variation mainly because of the correlation between the successive physical parameters $\bThe = \left(\bThe_1^T,\cdots,\bThe_M^T\right)^T$ (see Fig. \ref{fig:Real_Jason2_Echoes_JustifFilt}).  This smooth variation can be highlighted by expressing  the observed signals \eqref{eqt:Observation_model} as follows 
\begin{equation}
\bsy_k = \bss_k\left( \bThe \right) + \bse_k, \textrm{  with   } \bse_k \sim \calN \left(\bzero_M, \sigma^2_{k} \mathds{I}_M \right)
\label{eqt:Observation_model2}
\end{equation}
where $k\in\left\lbrace  1,\cdots,K\right\rbrace$ indexes the signal samples that are known as ``temporal gates'', $\mathds{I}_M$ denotes the ($M\times M$) identity matrix and $\bss_k$ is a smooth ($M\times 1$) vector representing the signal evolution at the $k$th gate (see Fig. \ref{fig:Real_Jason2_Echoes_JustifFilt} (bottom) for examples). The proposed Bayesian method aims to filter the observed signals $\bsy_k$, $k\in \{1,\cdots,K\}$, to retrieve the noiseless signals $\bss_k$, $k\in \{1,\cdots,K\}$.
The next section introduces the satellite altimetric model that will be considered in this paper since it satisfies the model described above. 

\subsection{Conventional altimetric model} \label{subsec:Conventional_model}
The altimetric model, in its simplified version, accounts for three parameters that are the amplitude $P_u$, the
epoch $\tau$ and the significant wave height $\textrm{SWH}$.
The resulting mathematical nonlinear model for the altimetric signal
is known as the ``Brown model'' and is given by \cite{Brown1977,Amarouche2004}
\begin{equation}
s(t)= \frac{P_u}{2}\left[ 1+ \textrm{erf} \left( \frac{t-\tau_s -\alpha
\sigma _{c}^{2}}{\sqrt{2}\sigma_{c}}\right) \right]
\exp\left[-\alpha \left( t-\tau_s  -\frac{\alpha \sigma
_{c}^{2}}{2}\right)
\right]   \label{eqt:Brown_model}
\end{equation}
where
\begin{equation}
\qquad \sigma _{c}^{2}=\left(\frac{\textrm{SWH}}{2c}
\right)^{2}+\sigma _{p}^{2}\label{modelecont2}
\end{equation} 
and where  $\textrm{erf}\left( t\right) =\frac{2}{\sqrt{\pi }}
\int_{0}^{t}e^{-z^{2}}\,dz$ stands for the Gaussian error
function, $t$ is the time, $\tau_s = \frac{2 \tau}{c}$ (resp. $\tau$) is the epoch expressed in seconds (resp. meters), $c$ is the speed of light, $\alpha$ and
$\sigma_{p}^{2}$ being two known parameters (depending on the
satellite and on the measurement instrument). The nonlinear model decribed in \eqref{eqt:Brown_model} is commonly used in the altimetric community mainly  because of its simplicity \cite{Amarouche2004,HalimiTGRS2012,HalimiTGRS2015}.
Note that the discrete altimetric signal is gathered in the vector $\bss = \left(s_1, \cdots, s_K\right)^T$, where $K=104$ gates, $s_k = s\left(k T \right)$,  $T$ is the time resolution and $\bThe_{m}=\left[\theta_1(m),\theta_2(m),\theta_3(m)\right] = \left[\textrm{SWH}(m),\tau(m),P_u(m)\right] $ is a ($1 \times 3$) vector containing the $3$ altimetric parameters $\textrm{SWH}, \tau, P_u$ for the $m$th signal.

The altimetric signals are corrupted by speckle noise that, thanks to the averaging that takes place on-board of the satellite, can be approximated by  additive Gaussian noise as shown in \cite{Puig2009,Germain2006,Halimi2013Eusipco,HalimiTGRS2015}. Thus, the observation altimetric model satisfies \eqref{eqt:Observation_model2}. Moreover, the  noise variances obtained, $\sigma^2_{k}, k\in\left\lbrace  1,\cdots,K\right\rbrace$, after the satellite averaging, are correlated due to the nature of the speckle noise (this correlation will be considered in the proposed Bayesian scheme).
Note that this paper only considers oceanic observations which generally show a smooth variation between successive signals. The next section introduces the Bayesian model associated with a set of $M$ successive signals considered in this paper.

\section{Hierarchical Bayesian model} \label{sec:Hierarchical_Bayesian_model}
This section introduces a hierarchical Bayesian model to denoise $M$ successive signals. The Bayesian approach first requires the determination of the likelihood that is based on the statistical model associated with the observed data. Second, the known properties of the parameters of interest are modeled via suitable prior distributions.  Bayes theorem allows the likelihood and the priors to be combined to build the posterior distribution of the statistical model. 
More precisely, if $f\left(\bsX \right)$  denotes the prior distribution assigned to the parameter $\bsX$, the Bayesian approach computes the posterior distribution of $\bsX$ using Bayes rule
\begin{equation}
f(\bsX|\bsY) = \frac{f(\bsY|\bsX) f(\bsX)}{f(\bsY)} \propto f(\bsY|\bsX) f(\bsX) \label{eqt:Bayes}
\end{equation}
where $\propto$ means ``proportional to'' and $f(\bsY|\bsX)$ is the likelihood of the observation vector $\bsY$.
The parameter $\bsX$ is then estimated from this posterior distribution by computing its mean (MMSE estimator) or its maximum (MAP estimator).
The following sections introduce the likelihood and the prior distributions considered in this paper. The unknown parameters of the proposed model include the ($K\times M$) matrix representing the noiseless signals $\bsS$, and the ($K \times 1$) vector $\bsig$ containing the noise variances associated with the  $M$ considered signals.

\subsection{Likelihood} \label{subsec:Likelihood}
The observation model defined in \eqref{eqt:Observation_model2} and the Gaussian properties of the noise sequence $\bse_k$, $k\in\left\lbrace  1,\cdots,K\right\rbrace$, yield
\begin{equation}
f(\bsy_{k}|\bss_k, \sigma^2_k) \propto {\left( \frac{1}{ \sigma^2_{k} }\right)}^{\frac{M}{2}} \exp
\left( -\frac{||\bsy_k - \bss_k ||^2}{2\sigma^2_k}   \right)  \label{eqt:likelihood}
\end{equation}
where  $ ||\cdot|| $ denotes the standard $l_2 $ norm such that $ ||\bsx||^2 = \bsx^T \bsx$ and $\bss_k\left( \bThe \right)$ has been denoted by $\bss_k$ for brevity. Assuming independence between the temporal samples of the observed signals leads to
\begin{equation}
f(\bsY| \bsS,\bThe) \propto \prod_{k=1}^{K}{f(\bsy_{k}|\bss_k, \sigma^2_k)}.
\label{eqt:likelihood_globale}
\end{equation}

\subsection{Priors for the observed signal} \label{subsec:Priors_for_the_observed_signal} 
As previously assumed, the successive observed signals evolve slowly leading to smooth vectors $\bss_k$, for $k\in{1,\cdots,K}$ (see Fig. \ref{fig:Real_Jason2_Echoes_JustifFilt} (bottom)).  This property is satisfied by considering a Gaussian prior for $\bss_k$ ensuring smoothness as follows
\begin{equation}
\bss_k |   \epsilon_k^2 \sim \calN \left(\bzero_{M}, \epsilon_k^2  \bsH \right),
\label{eqt:priorD}
\end{equation}
where $\bsH$ is an $(M\times M)$ matrix representing the squared-exponential covariance function given by  $H(m,m') = \exp\left[-\frac{\left(m-m'\right)^2}{\left(30\right)^2} \right]$, which introduces the correlation between the successive signals  and  $\epsilon_{k}^2$ is a variance parameter that is gate dependent. From \eqref{eqt:priorD}, it is clear that this variance is related to the energy of the signals at the $k$th gate (via the norm $\bss_{k}^T \bsH^{-1} \bss_{k}$). Moreover, because of the continuity of the signal $\bss_m$ w.r.t. the temporal gates, the signal energies vary smoothly from one gate to another. Therefore, we expect  $\epsilon_{k}^2$ to vary smoothly from one gate to another which will be introduced by considering a specific prior for $\epsilon_{k}^2 $, as explained in Section \ref{subsec:Hyperparameter_priors}.

\subsection{Prior for the noise variance} \label{subsec:Prior_for_the_Noise_variance} 
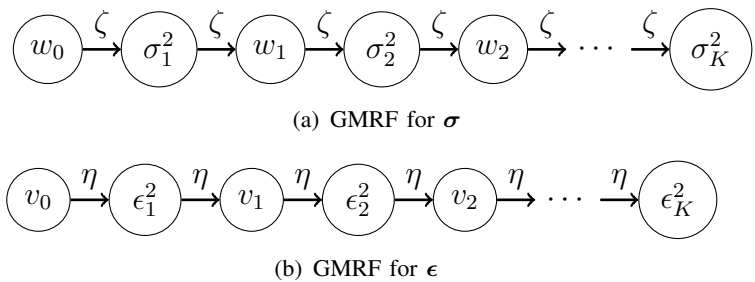
\begin{figure}
\centering
\subfigure[GMRF for $\bsig$]{
\begin{tikzpicture}
 nodes %
\node[draw, circle, text centered] (center) {$\sigma^2_2$}; 
\node[draw, circle,right =0.5  of center, text centered] (cr1) {$w_2$};
\node[right =0.5  of cr1, text centered] (cr2) {$\cdots$};
\node[draw, circle,right =0.5  of cr2, text centered] (cr3) {$\sigma^2_K$};
\node[draw, circle,left  =0.5  of center, text centered] (cl1) {$w_1$};
\node[draw, circle,left  =0.5  of cl1, text centered] (cl2) {$\sigma^2_1$};
\node[draw, circle,left  =0.5  of cl2, text centered] (cl3) {$w_0$};

\draw[->] (cl3) --(cl2) node[above,midway] {$\zeta$};
\draw[->] (cl2) --(cl1) node[above,midway] {$\zeta$};
\draw[->] (cl1) --(center) node[above,midway] {$\zeta$};
\draw[->] (center) --(cr1) node[above,midway] {$\zeta$};
\draw[->] (cr1) --(cr2) node[above,midway] {$\zeta$};
\draw[->] (cr2) --(cr3) node[above,midway] {$\zeta$};

 edges %
\draw[->, line width= 1] (cl3) to  [out=0,in=180, looseness=1] (cl2);
\draw[->, line width= 1] (cl2) to  [out=0,in=180, looseness=1] (cl1);
\draw[->, line width= 1] (cl1) to  [out=0,in=180, looseness=1] (center);
\draw[->, line width= 1] (center) to  [out=0,in=180, looseness=1] (cr1);
\draw[->, line width= 1] (cr1) to  [out=0,in=180, looseness=1] (cr2);
\draw[->, line width= 1] (cr2) to  [out=0,in=180, looseness=1] (cr3); 

\end{tikzpicture}}

\subfigure[GMRF for $\beps$]{
\begin{tikzpicture}
  nodes %
\node[draw, circle, text centered] (center) {$\epsilon^2_2$}; 
\node[draw, circle,right =0.5  of center, text centered] (cr1) {$v_2$};
\node[right =0.5  of cr1, text centered] (cr2) {$\cdots$};
\node[draw, circle,right =0.5  of cr2, text centered] (cr3) {$\epsilon^2_K$};
\node[draw, circle,left  =0.5  of center, text centered] (cl1) {$v_1$};
\node[draw, circle,left  =0.5  of cl1, text centered] (cl2) {$\epsilon^2_1$};
\node[draw, circle,left  =0.5  of cl2, text centered] (cl3) {$v_0$};

\draw[->] (cl3) --(cl2) node[above,midway] {$\eta$};
\draw[->] (cl2) --(cl1) node[above,midway] {$\eta$};
\draw[->] (cl1) --(center) node[above,midway] {$\eta$};
\draw[->] (center) --(cr1) node[above,midway] {$\eta$};
\draw[->] (cr1) --(cr2) node[above,midway] {$\eta$};
\draw[->] (cr2) --(cr3) node[above,midway] {$\eta$};

 edges %
\draw[->, line width= 1] (cl3) to  [out=0,in=180, looseness=1] (cl2);
\draw[->, line width= 1] (cl2) to  [out=0,in=180, looseness=1] (cl1);
\draw[->, line width= 1] (cl1) to  [out=0,in=180, looseness=1] (center);
\draw[->, line width= 1] (center) to  [out=0,in=180, looseness=1] (cr1);
\draw[->, line width= 1] (cr1) to  [out=0,in=180, looseness=1] (cr2);
\draw[->, line width= 1] (cr2) to  [out=0,in=180, looseness=1] (cr3); 

\end{tikzpicture}} \hspace{0.4cm}
\caption{Proposed 1st order GMRF neighborhood structures for (a) the noise variances $\bsig$  and (b) the signal energies $\beps$.}
\label{fig:GMRF}
\end{figure}

Due to the speckle origins of the corrupting noise, we expect the noise variances $\sigma^2_k$, $k\in\left\lbrace  1,\cdots,K\right\rbrace$ to vary smoothly. This behavior is considered by introducing an auxiliary vector $\bsw$ (of size  $K \times 1$) and  assigning a gamma Markov random field prior (GMRF) for the couple $(\bsig,\bsw)$ given by (see \cite{DikmenTASLP2010} for more details regarding this prior) 
\begin{eqnarray}
f\left(\bsig,\bsw | \zeta \right)  = &    \frac{1}{Z(\zeta)}  \left(\prod_{k=1}^{K} {\sigma_{k}^{-2(2\zeta+1)}} \right) \nonumber \\
\times & \left(\prod_{k'=1}^{K} {w_{k'}^{(2\zeta-1)}} \right)  \exp{\left(\frac{-\zeta w_0}{\sigma^2_1}\right)} \nonumber \\
\times & \prod_{k=1}^{K-1}  {\exp{\left[ -\zeta w_k \left(\frac{1}{\sigma^2_k}+ \frac{1}{\sigma^2_{k+1}}\right)  \right]}},
\label{eqt:priorEnergies}
\end{eqnarray}
where $Z(\zeta)$ is a normalizing constant  and  $\zeta>1$ is a fixed coupling parameter that controls the amount of correlation enforced by the GMRF. This prior ensures that each $\sigma^2_k$ is connected to two neighboring elements of $\bsw$ and vice-versa (see Fig. \ref{fig:GMRF} (a)).  Note that the variances $\sigma^2_k$ and $\sigma^2_{k'}$ for $k\neq k'$,  are conditionally independent and that the correlation is introduced via the auxiliary variables $\bsw$. An interesting property of this joint prior is that the conditional prior distributions of $\bsig$ and $\bsw$ reduce to conjugate inverse gamma ($\calI \calG$) and gamma ($\calG $) distributions, respectively, as follows \cite{DikmenTASLP2010}
\begin{eqnarray} 
\sigma_{k}^2 | w_{k-1},w_{k},\zeta & \sim &  \calI \calG \left[2\zeta, \zeta \left(w_{k-1}+w_{k}\right) \right]
\nonumber \\
\sigma_{K}^2 | w_{K-1},\zeta & \sim &  \calI \calG \left(\zeta, \zeta w_{K-1} \right)
\nonumber \\
w_{k}^2 | \sigma^2_{k},\sigma^2_{k+1},\zeta & \sim &    \calG \left[2\zeta, \frac{1}{\zeta \left(\frac{1}{\sigma^2_k}+ \frac{1}{\sigma^2_{k+1}}\right)}    \right]
\label{eqt:CondGam_IGamSig}
\end{eqnarray} 
where $k\in \left\lbrace1,\cdots,K-1\right\rbrace$.

\subsection{Hyperparameter priors} \label{subsec:Hyperparameter_priors}
As previously explained, the hyperparameters  $\epsilon_{k}^2$ are closely related to the signal energies via the norm $\left(\bss_{k}^T \bsH^{-1} \bss_{k}\right)$. Considering this property and the continuity of the signal suggest the presence of a correlation between the parameters $\epsilon_{k}^2$. This correlation can be introduced by considering a GMRF prior for $\left(\beps,\bsv\right)$ as follows \cite{DikmenTASLP2010}
\begin{eqnarray}
f\left(\beps,\bsv | \eta \right)  = &    \frac{1}{Z(\eta)}  \prod_{k=1}^{K} {\epsilon_{k}^{-2(2\eta+1)}} \nonumber \\
\times & \left(\prod_{k'=1}^{K} {v_{k'}^{(2\eta-1)}} \right)  \exp{\left(\frac{-\eta v_0}{\epsilon^2_1}\right)} \nonumber \\
\times & \prod_{k=1}^{K-1}  {\exp{\left[ -\eta v_k \left(\frac{1}{\epsilon^2_k}+ \frac{1}{\epsilon^2_{k+1}}\right)  \right]}},
\label{eqt:priorEnergiesEps}
\end{eqnarray}
where $\bsv$ are auxiliary variables and $\eta>1$ is the coupling parameter. A schematic description of the variable correlations is shown in Fig. \ref{fig:GMRF} (b)   which is similar to that presented in Section~\ref{subsec:Prior_for_the_Noise_variance}. The conjugate conditional prior distributions for  $\beps$ and $\bsv$ are given by 
\begin{eqnarray} 
\epsilon_{k}^2 | v_{k-1},v_{k},\eta  & \sim &   \calI \calG \left[2\eta, \eta \left(v_{k-1}+v_{k}\right) \right]
\nonumber \\
\epsilon_{K}^2 | v_{K-1},\eta  & \sim &   \calI \calG \left(\eta, \eta v_{K-1} \right)
\nonumber \\
v_{k}^2 | \epsilon^2_{k},\epsilon^2_{k+1},\eta  & \sim &     \calG \left[2\eta, \frac{1}{\eta \left(\frac{1}{\epsilon^2_k}+ \frac{1}{\epsilon^2_{k+1}}\right)}    \right]
\label{eqt:CondGam_IGamEps}
\end{eqnarray}
where $k\in \left\lbrace1,\cdots,K-1\right\rbrace$.

\subsection{Posterior distributions} \label{subsec:Posterior_distributions}
\begin{figure}
\centering
\begin{tikzpicture}
 nodes %
\node[text centered] (Y) {$\bsY$};
\node[above =1. of Y, text centered] (top1) {$ $};
\node[above =1. of top1, text centered] (top2) {$ $};
\node[above =1. of top2, text centered] (top3) {$ $};
\node[right  =0.4  of top1, text centered] (SIG) {$(\bsig,\bsw)$};
\node[left  =0.8  of top1, text centered] (X) {$\bsS$};
\node[draw, rectangle, right =2  of top2, text centered] (ZETA) {$\zeta$};
\node[left  =1.5  of top2, text centered] (EPS) {$(\beps,\bsv)$};
\node[draw, rectangle, left  =3  of top3, text centered] (ETA) {$\eta$};

edges %
\draw[->, line width= 1] (X) to  [out=270,in=145, looseness=0.5] (Y);
\draw[->, line width= 1] (SIG) to  [out=270,in=35, looseness=0.5] (Y); 
\draw[->, line width= 1] (EPS) to  [out=270,in=145, looseness=0.5] (X);
\draw[->, line width= 1] (ZETA) to  [out=270,in=35, looseness=0.5] (SIG);
\draw[->, line width= 1] (ETA) to  [out=270,in=145, looseness=0.5] (EPS);

\end{tikzpicture}
 
\caption{DAG for the parameter and hyperparameter priors (the user-fixed parameters appear in boxes). }
\label{fig:DAGs}
\end{figure}
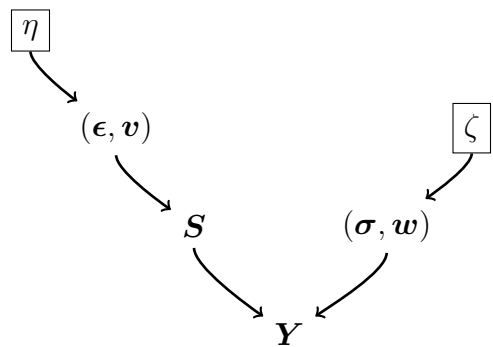

The proposed Bayesian model is summarized in the directed acyclic graph (DAG) displayed in Fig. \ref{fig:DAGs}. The parameters of interest are $\bsX = \left(\bsS,\bsig,\bsw,\beps,\bsv \right)$. The joint posterior distribution of this Bayesian model can be computed using the following hierarchical structure
\begin{equation}
f\left(\bsX | \bsY \right)  \propto f(\bsY|\bsS,\bsig)  f\left(\bsS |\beps  \right) f\left(\beps,\bsv\right) f\left(\bsig,\bsw\right),
\label{eqt:Joint_Posterior}
\end{equation}
where we have assumed a priori independence between the parameters. For simplicity,  $f(x|\theta)$ has been denoted  by $f(x)$ when the parameter $\theta$ is a user-fixed parameter. The MMSE and MAP estimators associated with the posterior \eqref{eqt:Joint_Posterior} are not easy to determine. In this paper, and akin to \cite{HalimiTGRS2015,HalimiArxiv2015b}, we propose to evaluate the MAP estimator by using an optimization technique maximizing the posterior \eqref{eqt:Joint_Posterior} w.r.t. the parameters of interest.

\section{Coordinate descent algorithm} \label{sec:Estimation_algorithms}
This section describes the optimization algorithm maximizing the posterior \eqref{eqt:Joint_Posterior} w.r.t. the noiseless signals and the noise variances. This provides the MAP estimator of the parameters of interest $\bsX$. An equivalent problem is to minimize w.r.t. $\bsX$, the negative log-posterior  $\mathcal{C}(\bsX)= - \textrm{log} [f(\bsX  | \bsY)]$  denoted as ``cost function'' and given by (after removing unnecessary constants)
\begin{eqnarray}
\mathcal{C} \left(\bsX  \right)   & =  & \sum_{k=1}^{K}{ \left[ \left(2\zeta + \frac{M}{2}+1\right)  \log \sigma_{k}^2  + 
\frac{\beta_{1}}{2\sigma_{k}^2} - \left(2\zeta-1\right) \log w_k \right.} \nonumber \\
& + & \left.\left(2\eta + \frac{M}{2}+1\right)  \log \epsilon_{k}^2 +
\frac{\beta_{2}}{2\epsilon_{k}^2} - \left(2\eta-1\right) \log v_k  \right]
\label{eqt:Cost_function}
\end{eqnarray}
where $\beta_{1} = ||\bsy_{k}-\bss_{k}||^2 + 2\zeta (w_{k-1}+w_{k})$ and 
$\beta_{2} = \bss_{k}^T  \bsH^{-1} \bss_{k}  + 2\eta (v_{k-1}+v_{k})$.
Because of the large number of parameters in $\bsX=\left(\bsS,\bsig,\bsw,\beps,\bsv \right)$, we propose a coordinate descent algorithm \cite{Bertsekas1995,Sigurdsson2014} that sequentially updates the different parameters. More precisely, in each step, the posterior distribution is maximized w.r.t. one parameter, the others being fixed. This process is repeated until the algorithm has converged to a local minimum of the cost function $\mathcal{C}(\bsS,\bsig,\bsw,\beps,\bsv)$. Thus, the algorithm iteratively updates each parameter by maximizing its conditional distribution as described in Algo. \ref{alg:Coordinate_descent_algorithm}. The next section describes the sub-optimization procedures maximizing the cost function $\mathcal{C}(\bsX)$ w.r.t. the noiseless signal $\bsS$, the noise variance $\bsig$ and the hyperparameters $\left(\bsw,\beps,\bsv \right)$.

\begin{algorithm}
\caption{Smooth Signal Estimation (SSE) Algorithm} \label{alg:Coordinate_descent_algorithm}
\begin{algorithmic}[1]
       \STATE \underline{Input}
       \STATE The noisy data $\bsY$, $zeta>1$, $eta>1$
       \STATE \underline{Initialization}
       \STATE Initialize parameters $\bsS^{(0)}$, $\bsig^{(0)}$, $\bsw^{(0)}$, $\bsv^{(0)}$, $\beps^{(0)}$ and $t=1$
       \STATE conv$=0$,
       \STATE \underline{Parameter update}
       \WHILE{conv$=0$}
               \STATE Update $\bss_{k}^{(t)},$ for $k \in  \left\lbrace 1,\cdots,K \right\rbrace$  according to \eqref{eqt:Fast_update_Sig}
               \STATE Update $\bsig^{(t)}$ according to \eqref{eqt:mode_sig}
						  	\STATE Update $\bsw^{(t)}$ according to \eqref{eqt:mode_w}
               \STATE Update $\beps^{(t)}$ according to \eqref{eqt:mode_eps}
               \STATE Update $\bsv^{(t)}$ according to \eqref{eqt:mode_v}
               \STATE Set conv$=1$ if the convergence criteria are satisfied
               \STATE $t = t + 1$
       \ENDWHILE 
\end{algorithmic}
\end{algorithm}

\subsubsection{Updating the parameters} \label{subsubsec:Updating_the_parameters}
The noiseless signal $\bsS$ can be updated by maximizing the conditional distribution associated with each independent $\bss_k$, which is a Gaussian distribution given by
\begin{equation}
\bss_k | \bsy_{k}, \sigma^2_{k}, \epsilon^2_{k}  \sim   \calN \left(\overline{\bss_k}, \bGam_{k} \right)
\label{eqt:posterior_mu_m}
\end{equation}
where
\begin{eqnarray} 
\overline{\bss_k}& = & \frac{1}{\sigma^2_{k}} \bGam_{k} \bsy_k, \label{eqt:posterior_mu_m_mean} \\
\bGam_{k} & = & \left(\frac{\bsH^{-1}}{\epsilon_k^2} + \frac{\mathds{I}_M}{\sigma^2_{k}}\right)^{-1}.   
\label{eqt:posterior_Sig_m_mean}
\end{eqnarray}
Therefore, the noiseless signal $\bsS$ can be updated using \eqref{eqt:posterior_mu_m_mean} which is the maximum of the Gaussian distribution. Note that this solution corresponds to a least squares solution of the quadratic problem w.r.t. $\bss_k$ shown in \eqref{eqt:Cost_function}. Note also that the matrix inversion in \eqref{eqt:posterior_mu_m_mean} should be computed at each descent step leading to a high computational cost. Thus, the proposed algorithm considers a useful modification to achieve this computation with less operations, as discussed in the Appendix.
The conditional distributions of $\bsig$, and  $\beps$ (resp. $\bsw$, and  $\bsv$) are inverse gamma distributions (resp. gamma distributions)  as follows 
\begin{eqnarray}
\sigma_{k}^2 | \bsy_{k}, \bss_{k}, \bsw_{k} \sim   \calI \calG \left(2\zeta + \frac{M}{2}, \frac{\beta_1}{2} \right) \\
\epsilon_{k}^2 | \bsy_{k}, \bss_{k}, \bsv_{k} \sim   \calI \calG \left(2\eta + \frac{M}{2}, \frac{\beta_2}{2} \right) \\
w_{k}^2 | \sigma^2_{k},\sigma^2_{k+1},\zeta \sim     \calG \left(2\zeta, \frac{1}{\zeta \left(\frac{1}{\sigma^2_k}+ \frac{1}{\sigma^2_{k+1}}\right)}    \right)\\
v_{k}^2 | \epsilon^2_{k},\epsilon^2_{k+1},\eta \sim     \calG \left(2\eta, \frac{1}{\eta \left(\frac{1}{\epsilon^2_k}+ \frac{1}{\epsilon^2_{k+1}}\right)}    \right).
\label{eqt:posterior_parameters}
\end{eqnarray}
The mode of each distribution is uniquely attained and given by 
\begin{eqnarray}
\overline{\sigma_{k}^2}   = \frac{\beta_1}{4\zeta + M +2} \label{eqt:mode_sig}\\
\overline{\epsilon_{k}^2} = \frac{\beta_2}{4\eta + M +2}  \label{eqt:mode_eps}\\ 
\overline{w_{k}^2} = 
\frac{2\zeta-1}{\zeta \left(\frac{1}{\sigma^2_k}+ \frac{1}{\sigma^2_{k+1}}\right)}  \label{eqt:mode_w}\\
 \overline{v_{k}^2} = \frac{2\eta-1}{\eta \left(\frac{1}{\epsilon^2_k}+ \frac{1}{\epsilon^2_{k+1}}\right)}.    \label{eqt:mode_v} 
\end{eqnarray}
These modes are used to update the parameters  $\bsig,\beps,\bsw,\bsv$ as shown in Algo. \ref{alg:Coordinate_descent_algorithm}. 

\subsubsection{Convergence and stopping criteria} \label{subsubsec:Convergence_and_stopping_criteria}
The coordinate descent algorithm converges to a stationary point of \eqref{eqt:Cost_function} provided that the minimum of that function w.r.t. $\bsX$ along each coordinate is unique (proposition 2.7.1  in  \cite{Bertsekas1995}). This is easily checked for all the parameters since they have unimode conditional distributions (Gaussian, gamma and inverse-gamma distributions).
The cost function is not convex, thus, the solution obtained  might depend on the initial values that should be chosen carefully.  In this paper, the parameters have been initialized as follows: $\bsig^{(0)}=\bss_m^{(0)}=\frac{1}{M} \sum_{n=1}^{M}{\bsy_n}$, $\forall m$,  $\epsilon_k^2=10$, $\forall k$,  and $w_k^{(0)} = v_k^{(0)}=10^{-12}$, $\forall k$. Note that more elaborate initialization procedures can be investigated, but these proposed values have provided relevant results in the considered simulations   (see Sections \ref{sec:Validation_on_synthetic_data} and \ref{sec:Results_on_Jason2_real_data}). 

Algo. \ref{alg:Coordinate_descent_algorithm} is an iterative algorithm that requires the definition of some stopping criteria. In this paper, we have considered two criteria and the algorithm is stopped if either of them is satisfied. The first criterion compares the new value of the cost function to the previous one and stops the algorithm if the relative error between these two values is smaller than a given threshold, i.e.,
\begin{equation}
| \mathcal{C} \left(\bsX^{t+1}\right)- \mathcal{C} \left(\bsX^{t}\right) | \leq  \xi \mathcal{C} \left(\bsX^{t}\right),
\label{eqt:criteria1}
\end{equation}
where $|.|$ denotes the absolute value and $\xi$ is the threshold that has been fixed to $\xi = 0.001$. The second criterion is based on a maximum number of iterations $T_{\textrm{max}}=100$.  The next sections study the behavior of the proposed algorithm when considering synthetic and real signals.

\section{Validation on synthetic data} \label{sec:Validation_on_synthetic_data}
This section evaluates the performance of the proposed algorithm with synthetic data. It is divided into two parts whose objectives are: 1) introducing the criteria used for the evaluation of the algorithm quality, 2) analyzing and comparing the behavior of the proposed algorithm with other state-of-the-art algorithms.

\subsection{Evaluation criteria} \label{subsec:Evaluation_criteria}
 
For synthetic signals, the quality of the proposed algorithm can be evaluated by comparing the noiseless signals $\bss_m$ to the denoised signals $\widehat{\bss}_m$ using the reconstruction signal to noise ratio (RSNR) given by \cite{BioucasWhispers2010}
\begin{equation}
 \textrm{RSNR} = 10 \log_{10} \left(\frac{\sum_{m=1}^{M}{||\bss_m||^2}}{    \sum_{m=1}^{M}{||\bss_m- \widehat{\bss}_m||^2}   } \right).
 \label{eqt:RSNR}
\end{equation}
Note that a high RSNR corresponds to a good denoising result.
Moreover, the true altimetric parameters are known for synthetic signals. Thus, the true values can be compared to the estimated ones before and after filtering to highlight the benefit of the proposed denoising algorithm. Note that the altimetric parameters have been estimated using the well known least-squares (LS) based strategy that is commonly used by the altimetric community \cite{Amarouche2004,HalimiTGRS2015}. The quality of the estimated parameters is evaluated using the root mean square error (RMSE) and the standard deviations (STDs) of the estimator $\widehat{\bthe_i}$ as follows 
\begin{eqnarray}
 \textrm{RMSE}\left( \widehat{\bthe_i}\right) & = &
\sqrt{\frac{1}{N}\sum_{n=1}^{N} \left[\widehat{\theta_i} {(n)}- \theta_i{(n)}
\right]^{2}},  
\label{eqt:RMSE} \\
 \textrm{STD}\left( \widehat{\bthe_i}\right)  & =  & \sqrt{
\frac{1}{N}\sum_{n=1}^{N} {\left[ \widehat{\theta_i} {(n)}
- \left(\frac{1}{N'}\sum_{n'=1}^{N'} \widehat{\theta_i} {(n')}\right) \right]}^2} \label{eqt:STD} 
\end{eqnarray}
for $i \in \left\lbrace1,\cdots,3 \right\rbrace$, where $\theta_i{(n)}$ (resp. $\widehat{\theta_i}{(n)}$ ) is the true (resp. estimated) parameter for the $n$th signal and $N$ is the number of simulated signals ($N'=N$ for synthetic signals).

When considering real signals, the performance of the proposed algorithm is qualitatively evaluated by a visual comparison between the noisy signals/parameters and the denoised ones \cite{Sandwell2005,Amarouche2004,HalimiTGRS2015}. Quantitatively, a modified parameter STD is computed using \eqref{eqt:STD} in which the averaged parameter value is approached by the mean of the estimated parameters along each $N'=20$ successive signals. This modified STD is called ``STD at 20 Hz'' \cite{Giles2012,Gommenginger2011OSTST,HalimiTGRS2014,HalimiTGRS2014b}.

\subsection{Simulation results on synthetic data} \label{subsec:Simulation_results_on_synthetic_data}

\begin{figure}[h!]
\centering \subfigure[SWH]{\includegraphics[width=0.75\figwidth,height=6cm]{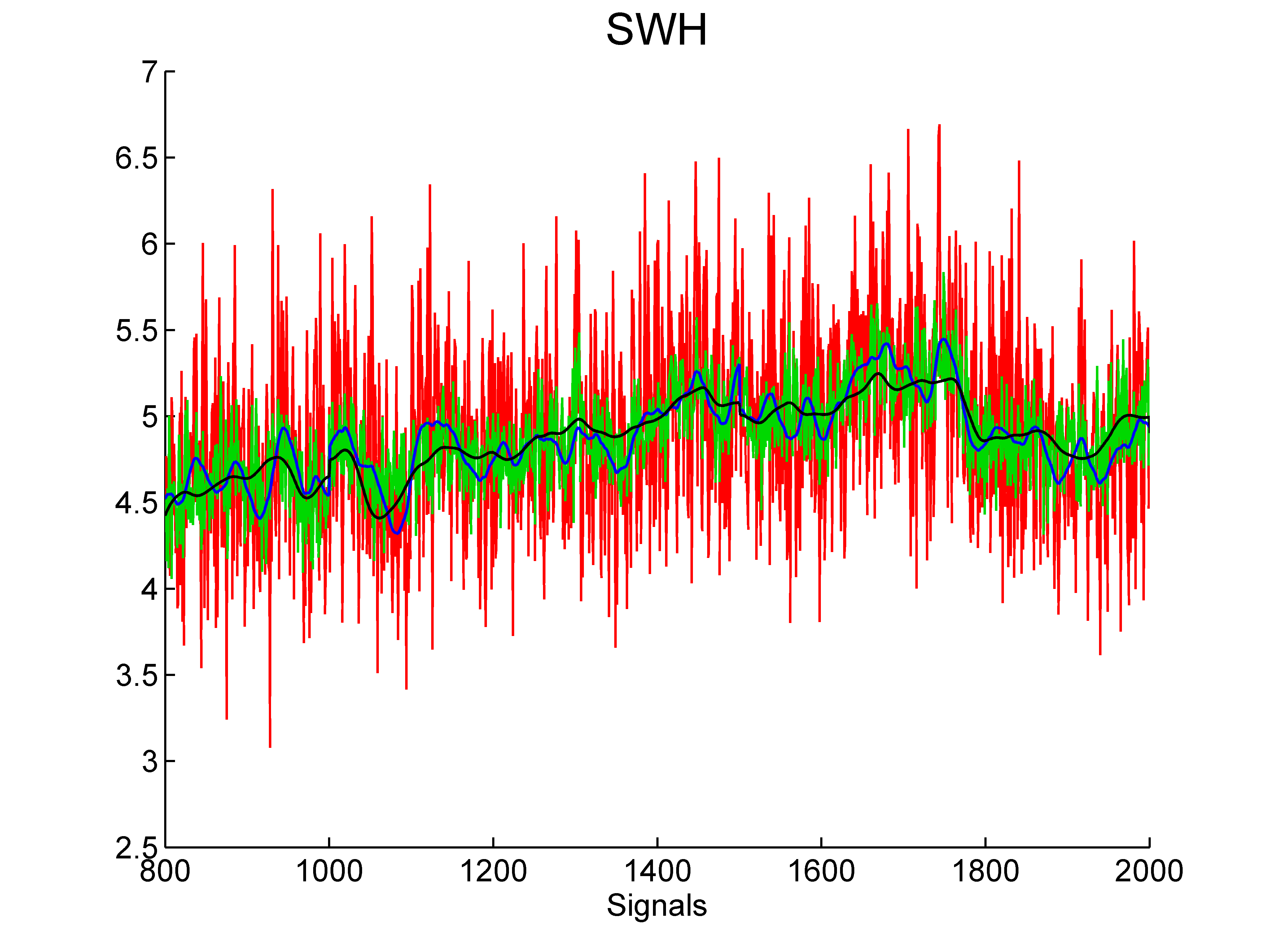}}
\subfigure[$\tau$]{\includegraphics[width=0.75\figwidth,height=6cm]{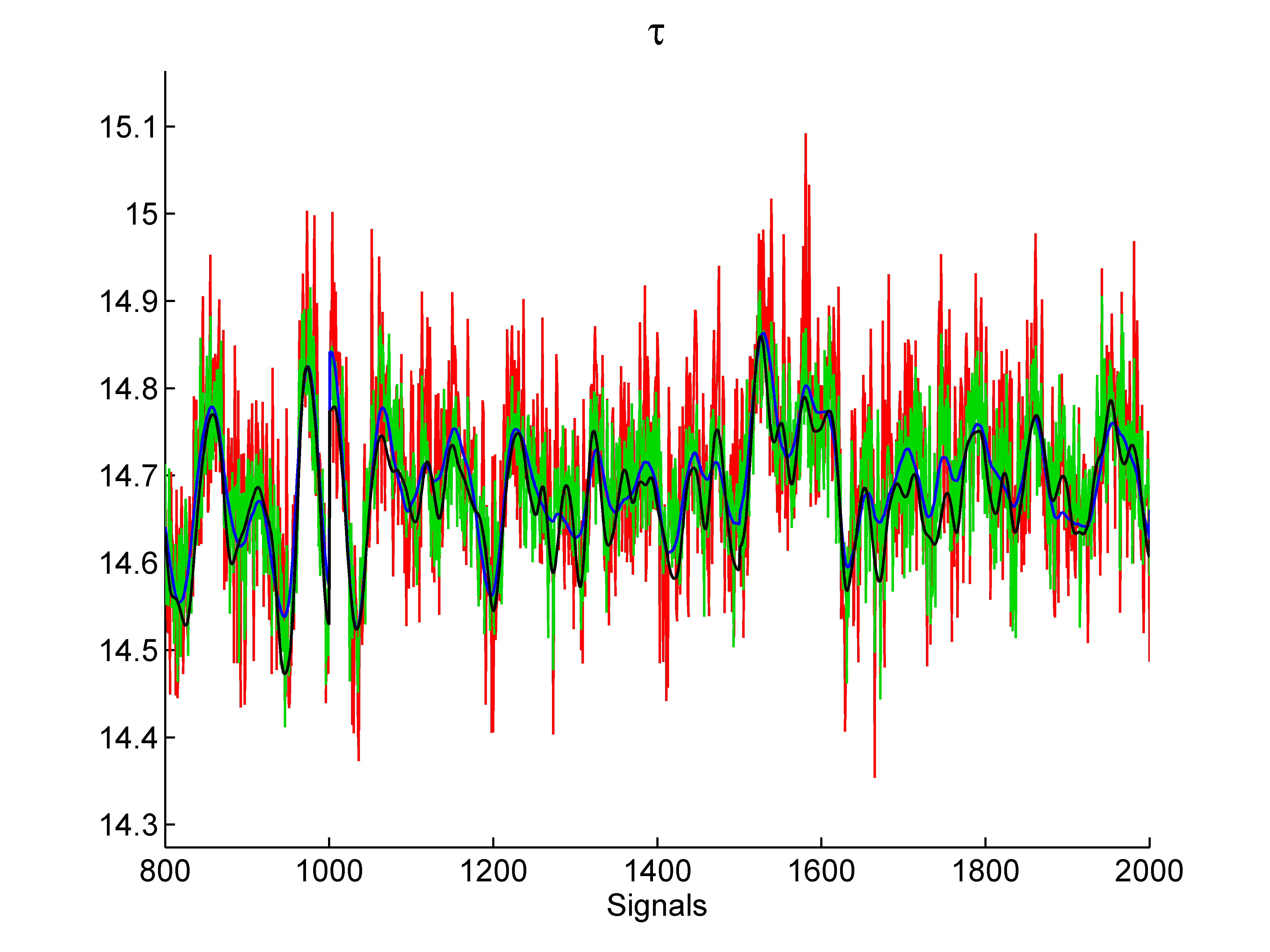}}
\subfigure[$P_u$]{\includegraphics[width=0.75\figwidth,height=6cm]{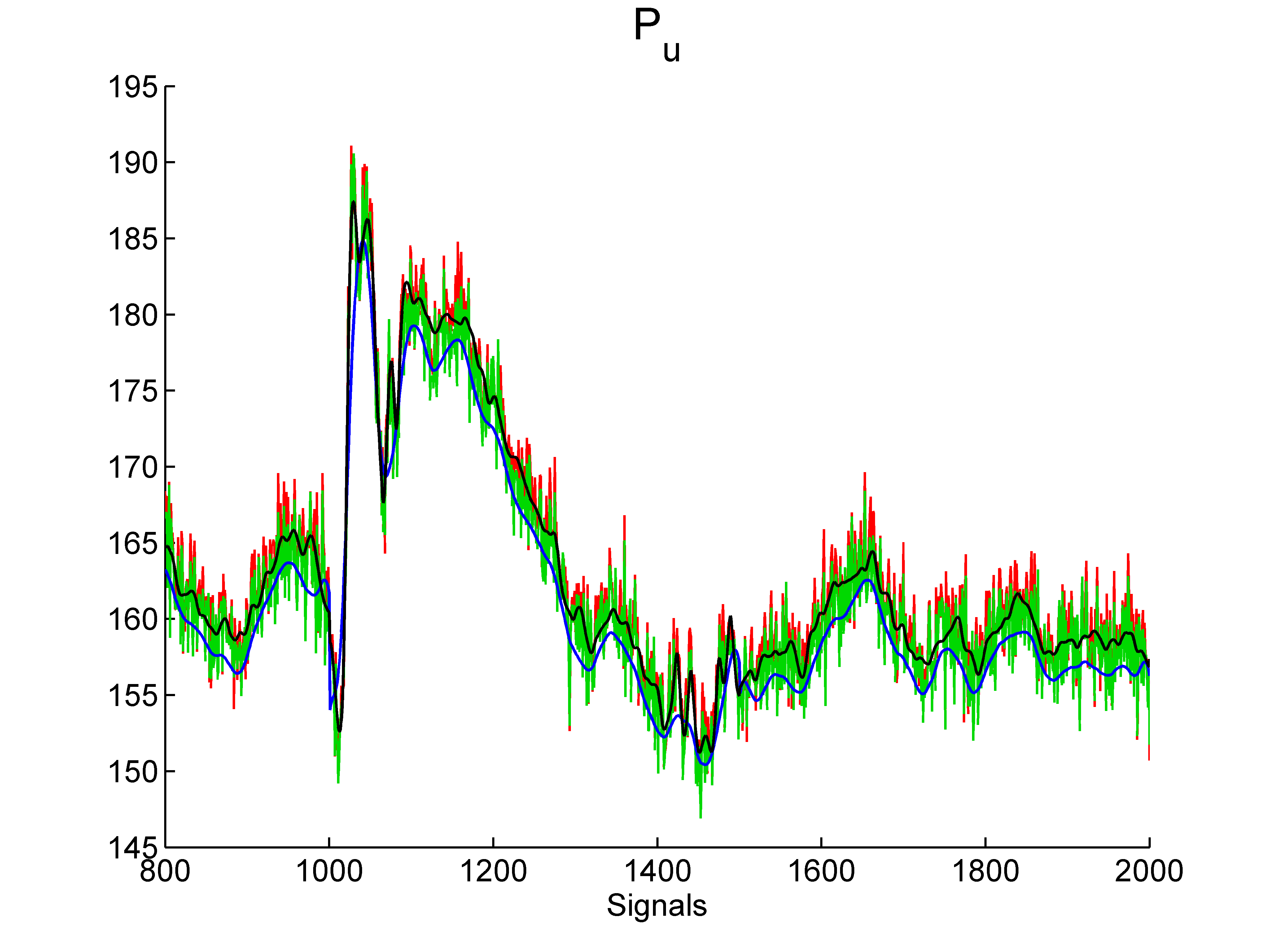}}
\caption{Example of $1200$ ground-truth synthetic parameters (black lines) and their estimations using the LS algorithm (red line), the SVD-LS algorithm with $M=500$ (green line)  and the proposed SSE-LS algorithm with $M=500$ (blue line). (a) SWH, (b) $\tau$ and (c) $P_u$. } \label{fig:Synth_evol_Param_Jason2_MLE_DMLE}
\end{figure}

Two experiments are conducted to evaluate the performance of the proposed SSE (for smooth signal estimation) algorithm. The first experiment studies the behavior of SSE when varying the number of the denoised signals $M$. Indeed, the SSE algorithm considers successive sets of length $M$ to denoise the observed altimetric signals. Therefore, $N=5000$ signals are generated according to the altimetric model \eqref{eqt:Brown_model} while using a realistic variation of the altimetric parameters  $\bThe_{m}= \left[\textrm{SWH}(m),\tau(m),P_u(m)\right]$. This realistic sequence of parameters is obtained by applying the CD-BM algorithm \cite{HalimiTGRS2015}, on $5000$ real Jason-$2$ signals (since it provides physically realistic smooth parameters) where we obtain $\textrm{SWH} \in [3.4,5.4]$m, $\tau \in [14.3,15]$ m and $P_u \in [150,190]$ unit (see Fig. \ref{fig:Synth_evol_Param_Jason2_MLE_DMLE} black lines).  The generated synthetic signals are then corrupted by speckle noise resulting from the averaging of $L=90$ signals and leading to $\textrm{RSNR}=19.55$ dB. The obtained $N=5000$ signals are processed by the proposed algorithm while considering different set lengths as shown in Table \ref{tab:Filter_length}. For example, for a length set $M=250$, the algorithm is run $20$ times to process the $N=5000$ signals. Overall, these results show an $\approx 11$ dB improvement in the processed data with an increasing RSNR w.r.t. $M$. However, a high number of $M$ requires higher computational cost (mainly due to the matrix inversion in \eqref{eqt:posterior_mu_m_mean}), while too small $M$ leads to more iterations. The value $M=500$ represents a good compromise and we consider this value for the rest of the paper \cite{HalimiTGRS2015}.  
 
\begin{table}[h] \centering
\centering \caption{Performance of the proposed SSE algorithm w.r.t. the filter length ($5000$ signals). The corrupted data presents an $\textrm{RSNR}= 19.55\textrm{\MakeLowercase{d}B}$.}
\begin{tabular}{|c|c|c|c|c|c|c|c|}
  \cline{2-8} \multicolumn{1}{c|}{} &  \multicolumn{7}{c|}{Filter length} \\
	\cline{2-8} \multicolumn{1}{c|}{} & 50 & 100 & 250 & 500 & 1000 & 2500 & 5000 \\
\hline RSNR (dB)   & 31.1 & 31.4 & 31.5 & 31.6 & 31.7 & 31.7 & 31.7 \\
\hline Time per signal (ms)  &  0.35 &  0.24 &  0.25 &  0.38 &  1.03 &  5.00 & 17.47
 \\
  \hline
\end{tabular}
\label{tab:Filter_length}
\end{table}

\begin{figure}[h!]
\centering
\includegraphics[width=0.75\figwidth]{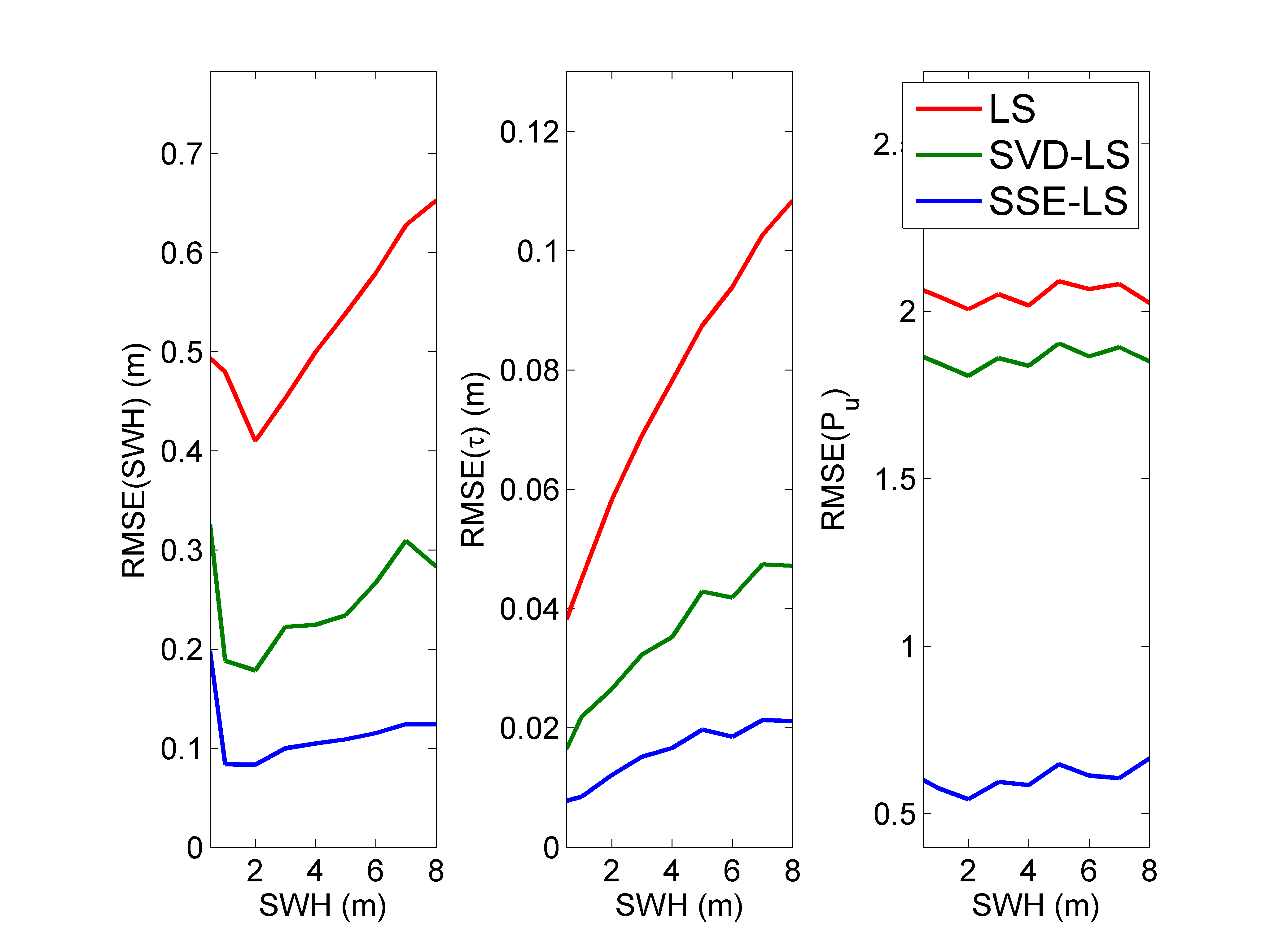}
\caption{RMSEs of the altimetric parameters w.r.t. SWH for LS, SVD-LS and SSE-LS algorithms.} \label{fig:RMSE_param_Fct_SWH_BM_MLE_MRFMLE}
\end{figure}

The second experiment studies the effect of the algorithm on the physical altimetric parameters $\left(\textrm{SWH},\tau,P_u\right)$. Indeed, it is of interest to devote more effort to improve the quality of the estimated altimetric parameters and to reduce the parameter standard-deviations \cite{Maus1998,Sandwell2005,HalimiTGRS2015,HalimiTGRS2014,HalimiTGRS2014b}. These parameters are estimated using the well known least-squares (LS)  based strategy \cite{Amarouche2004,HalimiTGRS2015} applied to noisy and filtered signals. The proposed strategy (denoted as SSE-LS) is compared to the classical LS algorithm  (without filtering) \cite{Amarouche2004,HalimiTGRS2015}, and to SVD-LS which is obtained by applying the singular value decomposition filtering strategy \cite{Ollivier2006,Thibaut2009OSTST} (with a threshold equal to 84$\%$) followed by the LS algorithm. Following \cite{HalimiTGRS2012,HalimiTGRS2014b}, the study is performed when varying SWH $\in  [0.5, 8]$ m  with fixed $\tau = 31$ gates and $P_u=130$. For each SWH, $500$ synthetic signals are generated using the Brown model with different noise realizations ($500$ Monte carlo runs) and processed using the three considered algorithms. Fig. \ref{fig:RMSE_param_Fct_SWH_BM_MLE_MRFMLE} presents the obtained parameter RMSEs w.r.t. SWH when considering the LS, SVD-LS and SSE-LS strategies. Overall, the proposed strategy presents the best performance with the lowest RMSEs. Indeed, at a typical SWH of $2$ m, the proposed SSE-LS reduces RMSE(SWH) to $10$ cm (against $40$ cm and $20$ cm  for LS and SVD-LS),  RMSE($\tau$) to $1$cm (against $6$ cm and $2.3$ cm  for LS and SVD-LS) and RMSE($P_u$) to $0.6$ (against $2$ and $1.8$ for LS and SVD-LS).  This parameter improvement is also highlighted in Fig. \ref{fig:Synth_evol_Param_Jason2_MLE_DMLE} which shows the estimated parameters when considering the first experiment settings (for clarity purpose, we only show $1200$ signal parameters). It is clear from this figure that SSE-LS (blue lines) provides smoother results that better approximate the actual parameters (black lines) than LS (red lines) and SVD-LS (green lines).
Table \ref{tab:PSNR_SWH} reports the obtained RSNR when considering SVD-LS and SSE-LS for different SWH. This table shows an average improvement by $6$ dB when considering SVD-LS and by $12$ dB when considering the proposed SSE-LS algorithm. It also shows a slightly decreasing performance for SVD-LS when increasing SWH while SSE-LS provides almost similar RSNR for all SWH. 
These results highlight the interest of the proposed strategy in denoising the altimetric signals and improving the estimated altimetric parameters. 

\begin{table}[h] \centering
\centering \caption{RSNR (in dB) with respect to SWH. The corrupted data presents an $\textrm{RSNR}= 19.55\textrm{dB}$.}
\begin{tabular}{|c|c|c|c|c|c|c|c|c|c|c|}
  \cline{3-11} \multicolumn{2}{c|}{} &  \multicolumn{9}{c|}{SWH (m)} \\
	\cline{3-11} \multicolumn{2}{c|}{} & 0.5 & 1 & 2 & 3  & 4  & 5  & 6  & 7  & 8  \\
\hline \multirow{2}{*}{RSNR (dB)}   & SVD-LS &  26.35 &  26.43 &  26.30 &  26.02 &  26.03 &  26.07 &  26.08 &  25.92 &  25.86 \\
\cline{2-11}                        & SSE-LS  &  32.24 &  32.21 &  32.22 &  32.13 &  32.15 &  32.10 &  32.22 &  32.13 &  32.07 \\
  \hline
\end{tabular}
\label{tab:PSNR_SWH}
\end{table}

\section{Results on Jason-$2$ real data}
\label{sec:Results_on_Jason2_real_data}
\begin{figure}[h!]
\centering
\includegraphics[width=0.75\figwidth]{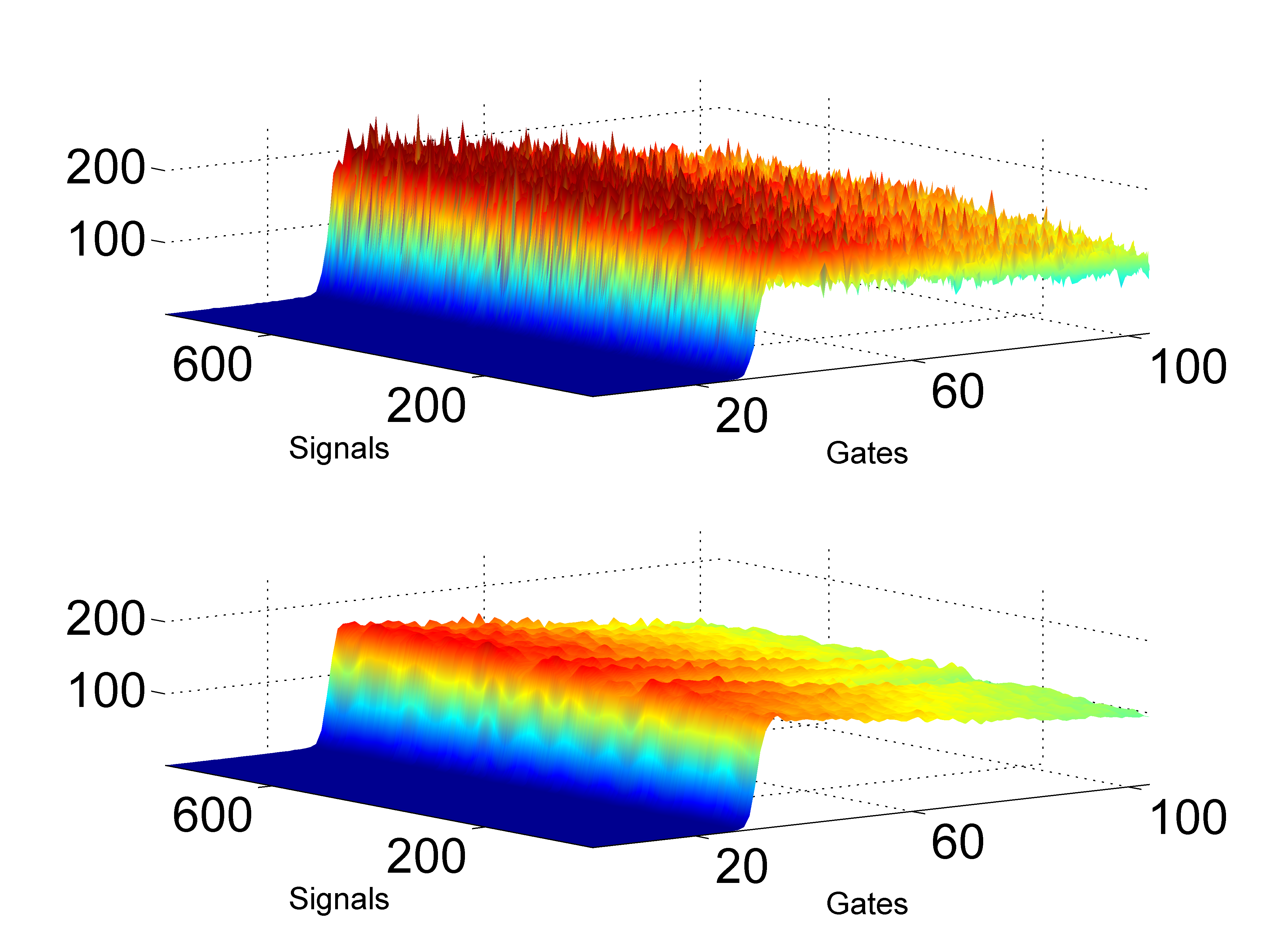}
\caption{Example of Jason-$2$ echoes. (top) without filtering and (bottom) with SSE filtering.} \label{fig:Real_Jason2_Echoes_MLE_DMLE}
\end{figure}

\begin{figure}[h!]
\centering \subfigure[SWH]{\includegraphics[width=0.75\figwidth,height=6cm]{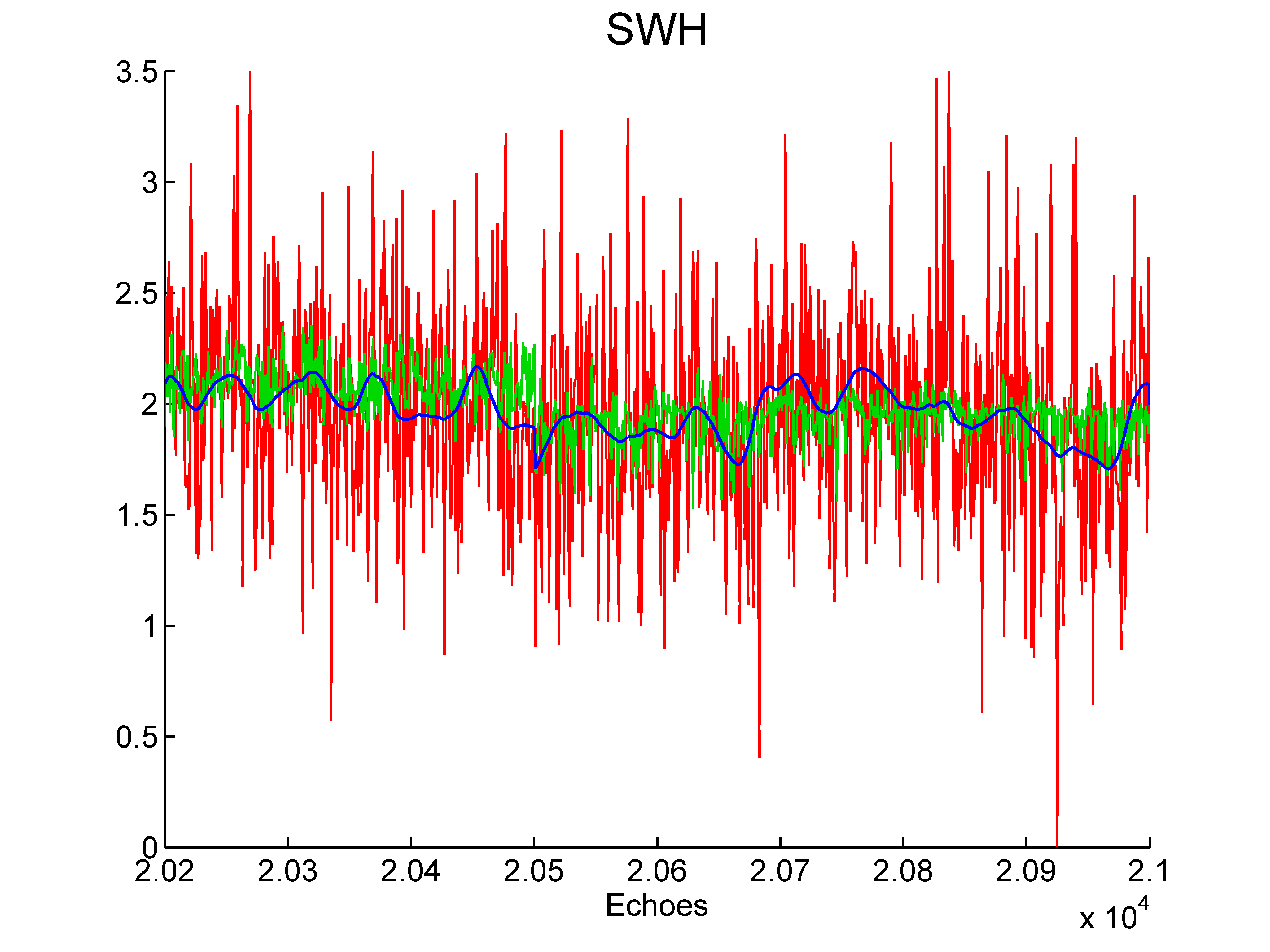}}
\subfigure[$\tau$]{\includegraphics[width=0.75\figwidth,height=6cm]{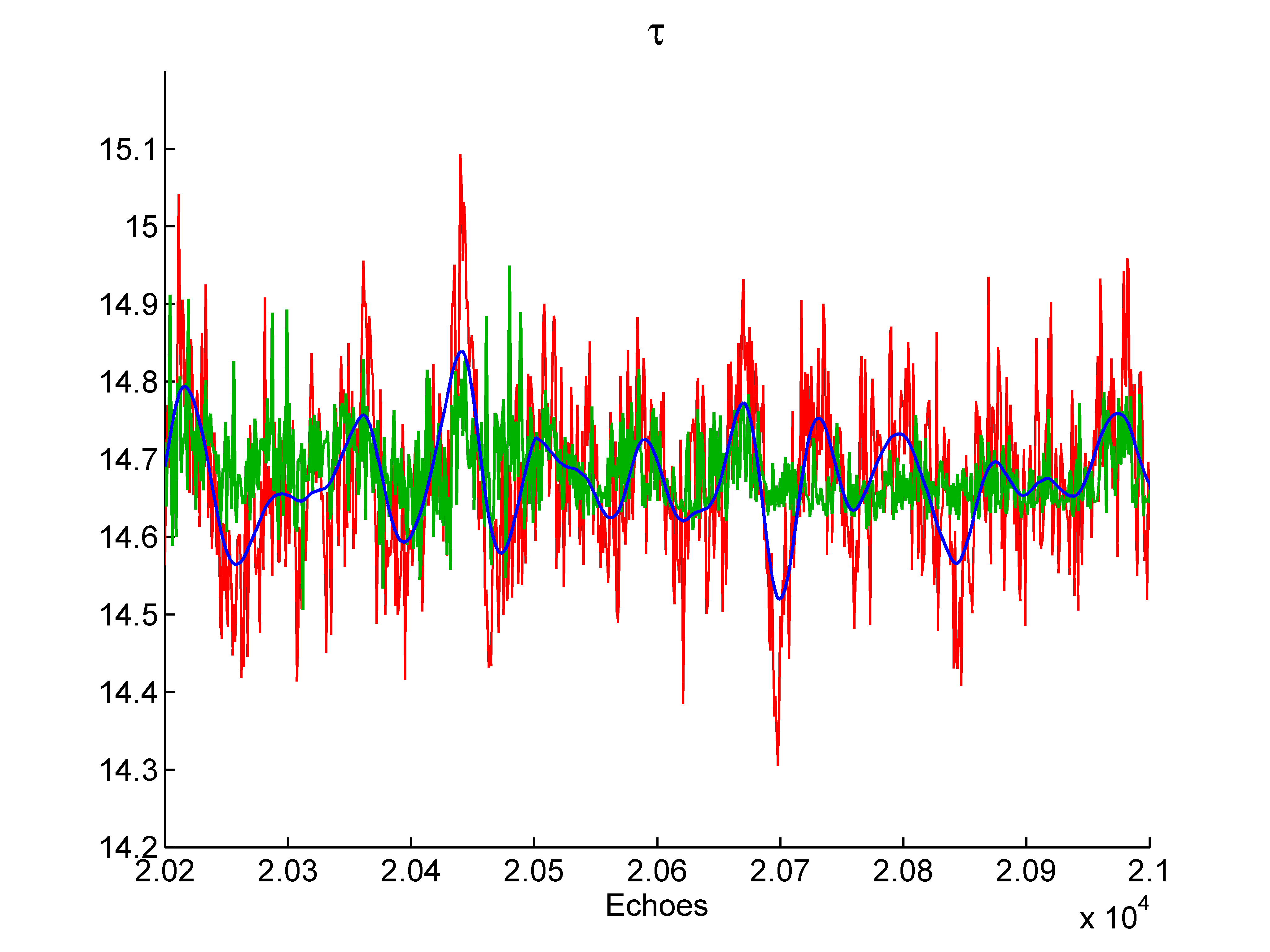}}
\subfigure[$P_u$]{\includegraphics[width=0.75\figwidth,height=6cm]{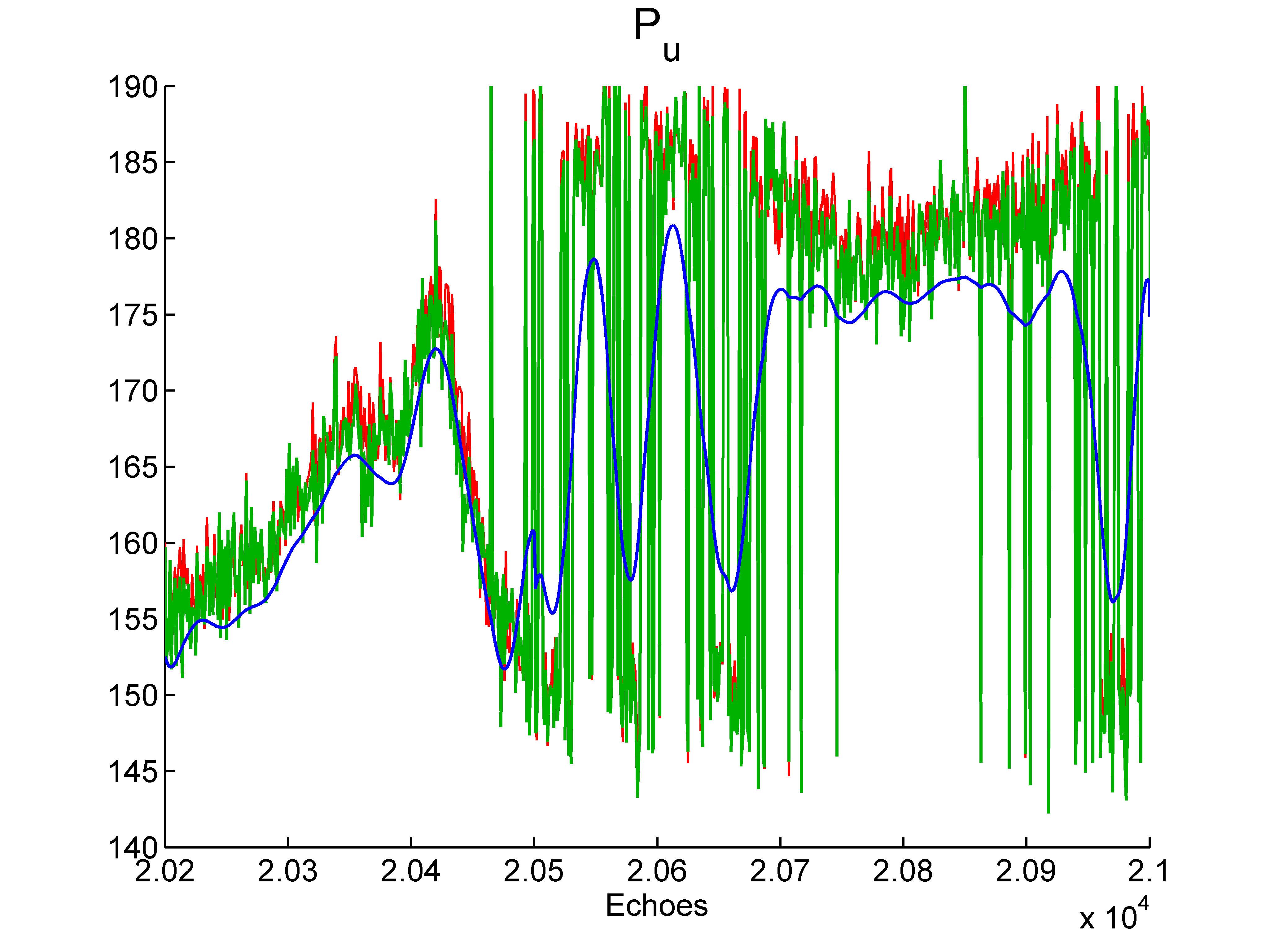}}
\caption{Estimated parameters using the LS algorithm (red line), the SVD-LS algorithm (green line)  and the proposed SSE-LS algorithm (blue line) for Jason-$2$ signals. (a) SWH, (b) $\tau$ and (c) $P_u$. } \label{fig:Real_evol_Param_Jason2_MLE_DMLE}
\end{figure}

This section is devoted to the validation of the proposed
SSE denoising algorithm when applied to the oceanic Jason-$2$ dataset.  The  data considered last for a period of $36$ minutes and consist  of $43000$ real signals that were extracted from pass $30$ of cycle $35$.  Fig. \ref{fig:Real_Jason2_Echoes_MLE_DMLE} presents a sequence of $800$ Jason-$2$ signals before and after filtering. Note first that this sequence  shows a reduced variation in the altimetric signals which justifies the use of the proposed strategy.
Moreover, it clearly shows a reduction in the noise affecting the signals after the application of the SSE algorithm especially in the tail of the signal (the decreasing part), which was most affected by the speckle noise. Fig. \ref{fig:Real_evol_Param_Jason2_MLE_DMLE}  shows the parameters estimated  w.r.t. time when considering the  LS (in red), SVD-LS (in green) and SSE-LS (in blue) algorithms. As observed for synthetic data in Section \ref{subsec:Simulation_results_on_synthetic_data}, the proposed SSE-LS provides a smooth parameter evolution which is physically more consistent, while SVD-LS and LS present high estimation noise (a lot of vibrations especially for $P_u$). 
This result is quantitatively confirmed in Table \ref{tab:Bias_STD_Time_Real} which shows smaller STDs for SSE-LS than for LS and SVD-LS. This STD reduction is of great importance for many practical applications related to oceanography such as bathymetry. Comparing SSE-LS to LS, Table \ref{tab:Bias_STD_Time_Real} highlights an STD improvement factor by $6$ for SWH, $4$ for $\tau$ and $5$ for $P_u$. This table also shows a good agreement between the means of the estimated parameters for the LS, SVD-LS and SSE-LS algorithms (except $P_u$ that is slightly reduced by SSE-LS as shown in Fig. \ref{fig:Real_evol_Param_Jason2_MLE_DMLE}). 
Finally, Table \ref{tab:Bias_STD_Time_Real} also compares the computational costs of the three considered algorithms when processing the $43000$ signals (the result is reported for each signal). Because of the filtering step, both SVD-LS and SSE-LS require more computational times than LS. Note that the proposed SSE algorithm requires more computational time than the SVD approach. However, this cost (about $12\%$ of additional computational times w.r.t. the LS algorithm) must be balanced by the performance improvement in terms of RSNR and parameter STDs. These results confirm the good performance of the proposed strategy for denoising smooth signals such as oceanic altimetric signals.

\begin{table}[h] \centering
\centering \caption{Performance on real Jason-$2$ data ($45000$ signals).}
\begin{tabular}{|c|c|c|c|c|}
  \cline{3-5}
\multicolumn{2}{c|}{}                        & SWH (cm)   & $\tau$ (cm) & $P_u$ \\
\hline  \multirow{3}{*}{Mean}   & LS   & 242      & 14.68     & 167.73 \\
\cline{2-5}                     & SVD-LS  & 241      & 14.67     & 166.62 \\
\cline{2-5}                     & SSE-LS  & 248      & 14.68     & 164.83 \\
\hline
\hline \multirow{3}{*}{STD}     & LS   & 59.9    & 12.01   & 6.18 \\
\cline{2-5}                     & SVD-LS  & 18.14   & 6.02   & 6.09 \\
\cline{2-5}                     & SSE-LS  & \textbf{9.03}   & \textbf{2.94}   & \textbf{1.21}  \\
\hline
\hline \multirow{2}{*}{Average time}& LS  & \multicolumn{3}{c|}{\textbf{8.56}}  \\
\cline{2-5} \multirow{2}{*}{per signal (ms) }  & SVD-LS  & \multicolumn{3}{c|}{9.05}  \\
\cline{2-5}    & SSE-LS  & \multicolumn{3}{c|}{9.63}  \\
  \hline
\end{tabular}
\label{tab:Bias_STD_Time_Real}
\end{table}

\section{Conclusions} \label{sec:Conclusions}
This paper has presented a new Bayesian strategy for the estimation of smooth signals corrupted by Gaussian noise. The successive continuous signals can have a numerical expression or be given by a linear/nonlinear function with respect to some parameters. A Bayesian model was proposed to take into account the Gaussian properties of the noise and the smooth properties of the signal evolutions. Moreover, the signal energies were assigned a GMRF prior that introduces correlation between their values to account for their continuity. Similarly, the noise variances were also assigned a GMRF to better approximate the speckle noise that can affect the signals. The resulting posterior distribution was maximized using a fast coordinate descent algorithm that showed good results on both synthetic and real altimetric signals. The proposed algorithm was also evaluated by combining it with a commonly used parameter estimation strategy for the altimetric parameters. The estimated parameters showed a clear improvement highlighting the benefit of the proposed algorithm. It is worth-noting that the proposed strategy is fast and generic and thus could be applied when considering other altimetric technologies such as delay/Doppler altimetry \cite{Raney1998,HalimiTGRS2014,HalimiTGRS2014b}. This point will be considered in future work. Generalizing the proposed approach for hyperspectral images is also an interesting issue that is currently under investigation.

\clearpage
\appendix[Mathematical derivations] \label{app:Mathematical_derivations}

\subsection{Updating the noiseless signal $\bss_k$} \label{app:Matrix inversion computation}
The ($M\times M$) matrix inversion in \eqref{eqt:posterior_Sig_m_mean} should be computed at each update of the noiseless signals which requires a high computational cost. To avoid this cost, we divide this matrix inversion into two parts. One representing the heavy computations and is achieved outside the ``while'' loop in Algo. \ref{alg:Coordinate_descent_algorithm}. The other one includes simple vector multiplications and is kept inside the loop. To achieve this, an SVD decomposition is first applied to $\bsH^{-1}$ as follows
\begin{equation}
\bsH^{-1} = \bsV  \bsD\left(r_i\right) \bsV^T 
\label{eqt:SVD_H}
\end{equation}
where $\bsD\left(x_i\right)$ denotes a diagonal matrix with its $i$th diagonal element equal to $x_i$, $r_i$ is the $i$th singular value of $\bsH^{-1}$ and $\bsV$ is a unitary orthogonal matrix, i.e., $\bsV \bsV^T =\mathds{I}_M$. Straightforward computations lead to the following expression for the noiseless signal update  
\begin{equation}
\overline{\bss_k} = \bsV   \bsD\left(\frac{\epsilon_k^2}{ r_i \sigma^2_k + \epsilon_k^2 }\right)  \underline{\bsV^T \bsy_k}. 
\label{eqt:Fast_update_Sig}
\end{equation}
Note that the operation underlined  in \eqref{eqt:Fast_update_Sig} and the SVD decomposition \eqref{eqt:SVD_H} are only computed once  outside the loop while the remaining vector operations in \eqref{eqt:Fast_update_Sig} are achieved inside the loop.

\newpage
\bibliographystyle{ieeetran} 
\bibliography{biblio_all}

\end{document}

%% file: notations.tex
\input com_notations.tex

\newcounter{algo}
\renewcommand{\thealgo}{\arabic{algo}}

%% file: com_notations.tex
\def\bse{{\boldsymbol{e}}}

\def\bss{{\boldsymbol{s}}}

\def\bsv{{\boldsymbol{v}}}
\def\bsw{{\boldsymbol{w}}}
\def\bsx{{\boldsymbol{x}}}
\def\bsy{{\boldsymbol{y}}}

\def\bsD{{\boldsymbol{D}}}

\def\bsH{{\boldsymbol{H}}}

\def\bsS{{\boldsymbol{S}}}

\def\bsV{{\boldsymbol{V}}}

\def\bsX{{\boldsymbol{X}}}
\def\bsY{{\boldsymbol{Y}}}

\def\calG{{\mathcal{G}}}

\def\calI{{\mathcal{I}}}

\def\calN{{\mathcal{N}}}